%% file: main.tex
\documentclass[fleqn,usenatbib]{mnras}

\usepackage{newtxtext,newtxmath}
\usepackage[T1]{fontenc}

\DeclareRobustCommand{\VAN}[3]{#2}
\let\VANthebibliography\thebibliography
\def\thebibliography{\DeclareRobustCommand{\VAN}[3]{##3}\VANthebibliography}

\usepackage{graphicx}	
\usepackage{amsmath}	
\title[ASKAP follow-up of GW190814]{A comprehensive search for the radio counterpart of GW190814 with the Australian Square Kilometre Array Pathfinder}

\author[D. Dobie et al.]{D. Dobie$^{1,2}$,
A. Stewart$^{3}$,
K. Hotokezaka$^{4}$,
Tara Murphy$^{3,2}$,
D.~L.\ Kaplan$^{5}$,
D.A.H. Buckley$^{6,7,8}$,
J. Cooke$^{1,2}$,
\newauthor
A. Y. Q. Ho$^{9,10}$,
E. Lenc$^{11}$,
J. K. Leung$^{3,11,2}$
M. Gromadzki$^{12}$
A. O'Brien$^{5}$,
S. Pintaldi$^{13}$,
J. Pritchard$^{3,11,2}$
\newauthor
Y. Wang$^{3,11,2}$
Z. Wang$^{3,11,2}$
\\
$^{1}$Centre for Astrophysics and Supercomputing, Swinburne University of Technology, Hawthorn, Victoria, Australia\\
$^{2}$ARC Centre of Excellence for Gravitational Wave Discovery (OzGrav), Hawthorn, VIC 3122, Australia\\
$^{3}$Sydney Institute for Astronomy, School of Physics, The University of Sydney, NSW 2006, Australia\\
$^{4}$Research Center for the Early Universe, Graduate School of Science, University of Tokyo, Bunkyo, Tokyo 113-0033, Japan\\
$^{5}$Department of Physics, University of Wisconsin-Milwaukee, P.O. Box 413, Milwaukee, WI 53201, USA\\
$^{6}$South African Astronomical Observatory, PO Box 9, Observatory Road, Observatory 7935, Cape Town, South Africa\\
$^{7}$Department of Astronomy, University of Cape Town, Private Bag X3, Rondebosch 7701, South Africa\\
$^{8}$Department of Physics, University of the Free State, PO Box 339, Bloemfontein 9300, South Africa\\
$^{9}$Department of Astronomy, University of California, Berkeley, 501 Campbell Hall, Berkeley, CA, 94720, USA\\
$^{10}$Miller Institute for Basic Research in Science, 468 Donner Lab, Berkeley, CA 94720, USA\\
$^{11}$CSIRO, Space and Astronomy, PO Box 76, Epping, NSW 1710, Australia\\
$^{12}$Astronomical Observatory, University of Warsaw, Al. Ujazdowskie 4, 00-478 Warszawa, Poland\\
$^{13}$Sydney Informatics Hub, The University of Sydney, NSW 2008, Australia
}

\date{Accepted XXX. Received YYY; in original form ZZZ}

\pubyear{2021}

\begin{document}
\label{firstpage}
\pagerange{\pageref{firstpage}--\pageref{lastpage}}
\maketitle

\begin{abstract}
We present results from a search for the radio counterpart to the possible neutron star--black hole merger GW190814 with the Australian Square Kilometre Array Pathfinder. We have carried out 10 epochs of observation spanning 2--655 days post-merger at a frequency of 944\,MHz. Each observation covered 30\,deg$^2$, equivalent to 87\% of the event localisation. We conducted an untargeted search for radio transients in the field, as well as a targeted search for transients associated with known galaxies. We find one radio transient, ASKAP~J005022.3$-$230349, but conclude that it is unlikely to be associated with the merger. We use our observations to place constraints on the inclination angle of the merger and the density of the surrounding environment by comparing our non-detection to model predictions for radio emission from compact binary coalescences. This survey is also the most comprehensive widefield search (in terms of sensitivity and both areal and temporal coverage) for radio transients to-date and we calculate the radio transient surface density at 944\,MHz.
\end{abstract}

\begin{keywords}
radio continuum: transients -- gravitational waves -- black hole - neutron star mergers
\end{keywords}



\section{Introduction}
The detection of light and gravitational waves from a neutron star merger, GW170817, had profound implications for astrophysics \citep[][]{2017PhRvL.119p1101A,2017ApJ...848L..12A}. While there were significant results from the thermal (``kilonova'') emission seen in the optical and near-infrared \citep[e.g.]{2017ApJ...848L..27T,2017Natur.551...67P,2017ApJ...848L..17C,2017Natur.551...80K,2017Sci...358.1559K} observations of the synchrotron after were uniquely able to shed light on  the nature of the jet launched by the merger and allowed measurements of the total energy released, the circum-merger density and the merger's inclination angle \citep[e.g.][]{2018Natur.561..355M,2019Sci...363..968G,2021arXiv210402070H,2020arXiv200602382M,2020MNRAS.498.5643T}. While the radio counterpart to GW170817 was only discovered as part of targeted observations of the optical counterpart \citep[][]{2017Sci...358.1579H}, it would have been possible to discover it (albeit at a much later time) as part of a radio-only search \citep[][]{2021MNRAS.505.2647D}. Similarly, we expect that some future mergers will be detectable with radio observations alone, although this is dependent on the typical properties of these afterglows including the jet structure, energetics and microphysics parameters. Here we illustrate future capabilities by presenting a case study of the search for radio emission from the potential neutron star-black hole merger GW190814.

GW190814 \citep{2020ApJ...896L..44A} is a compact binary coalescence detected during the third LIGO/Virgo observing run (O3) at a distance of $241^{+41}_{-45}$\,Mpc. The primary component is a black hole with mass $23.2^{+1.1}_{-1.0}$\,M$_{\sun}$, and the secondary component has mass $2.59^{+0.08}_{-0.09}$\,M$_{\sun}$. While current understanding suggests that this component was a black hole \citep[e.g.][]{2020ApJ...904...80E,2021ApJ...908L..28N,2021ApJ...908L...1T}, a neutron star cannot be completely ruled out \citep[e.g.][]{2021MNRAS.505.1600B,2021ApJ...908..122G,2021Ap&SS.366....9R,2021ApJ...910...62Z}. 

Despite a comprehensive multi-wavelength follow-up effort, no counterparts have been detected in optical/near-IR \citep{2019ApJ...884L..55G,2020A&A...643A.113A,2020ApJ...890..131A,2020MNRAS.492.3904A,2020MNRAS.497..726G,2020ApJ...901...83M,2020MNRAS.499.3868T,2020ApJ...895...96V,2020MNRAS.492.5916W,2021arXiv210302399D,2021arXiv210606897K}, radio \citep{2019ApJ...887L..13D,2021arXiv210208957A} or X-ray \citep[][Cenko et al. in prep.]{2020MNRAS.499.3459P} observations. Even if the secondary component is a neutron star it is unclear whether this event would be expected to produce an electromagnetic counterpart as current neutron star--black hole (NSBH) models make a large range of predictions spanning no counterpart at any wavelength, to kilonovae that are an order of magnitude brighter than neutron star mergers \citep[e.g.][]{2016ApJ...831..190H,2017Natur.551...80K,2017CQGra..34j4001R,2021arXiv210615781Z}. However, for GW190814 in particular, the extreme mass ratio \citep[$0.112_{-0.009}^{+0.008}$;][]{2020ApJ...896L..44A} suggests that the merger likely immediately formed a black hole, ruling out the presence of any counterparts associated with the collapse of a supramassive neutron star into a black hole, although this does not preclude the presence of radio emission produced by relativistic tidal debris \citep[e.g.][]{2011Natur.478...82N}.

Several events detected during O3 had southern localisations with at least one plausible neutron star component, making follow-up with the Australian Square Kilometre Array Pathfinder \citep[][]{2021PASA...38....9H} feasible. However, most were localised to hundreds of square degrees and while ASKAP is capable of following up these events \citep{2019PASA...36...19D}, we instead focused on events that could be followed up with a single pointing. GW190814 was initially localised to 23\,$\deg^2$, which has since improved to 18.5\,$\deg^2$, and was the the only event suitable for single-pointing ASKAP follow-up\footnote{We also performed observations of the candidate binary neutron star merger S190510g, but this event was later reclassified as having a terrestrial origin.}. Our observations covered 89\% of the initial localisation area within a single pointing, centered on the skymap posterior maximum.

In \citet{2019ApJ...887L..13D} we reported an initial search for a counterpart spanning 2--33 days post-merger. While no counterpart was detected, these observations allowed us to rule out the presence of an on-axis relativistic jet which would have been detectable at the early times we observed. However, the final gravitational wave parameter estimates for this event later suggested the inclination angle was $\theta_{\rm obs}=46^{+17}_{-11}\,\deg$ meaning that the radio lightcurve would be expected to peak at much later times. The potential delay between the merger, the optical emission and the radio afterglow is not unexpected and has been observed previously in gamma-ray bursts and other transient classes, where the GHz-frequency radio emission may not peak until months post-event \citep[][]{2021NatAs...5..491H,2021MNRAS.503.1847L}.

In this paper we report the results of our continued monitoring with ASKAP out to 655 days post-merger. In Section \ref{sec:data} we outline the observations that were carried out and describe the details of our search for transient and variable sources. In Section \ref{sec:discussion} we discuss the candidate counterparts found in our search and ultimately rule them all out as related to GW190814 based on their lightcurve morphology, archival radio data from the Karl G. Jansky Very Large Array (VLA) and other multi-wavelength observations. We also discuss the constraints that our observations place on the properties of any jet launched by the merger, the implications of this search for the field of radio transient astronomy and evaluate the various follow-up strategies that have been applied to this event.

\input{obs_table}

\begin{figure*}
    \centering
    \includegraphics{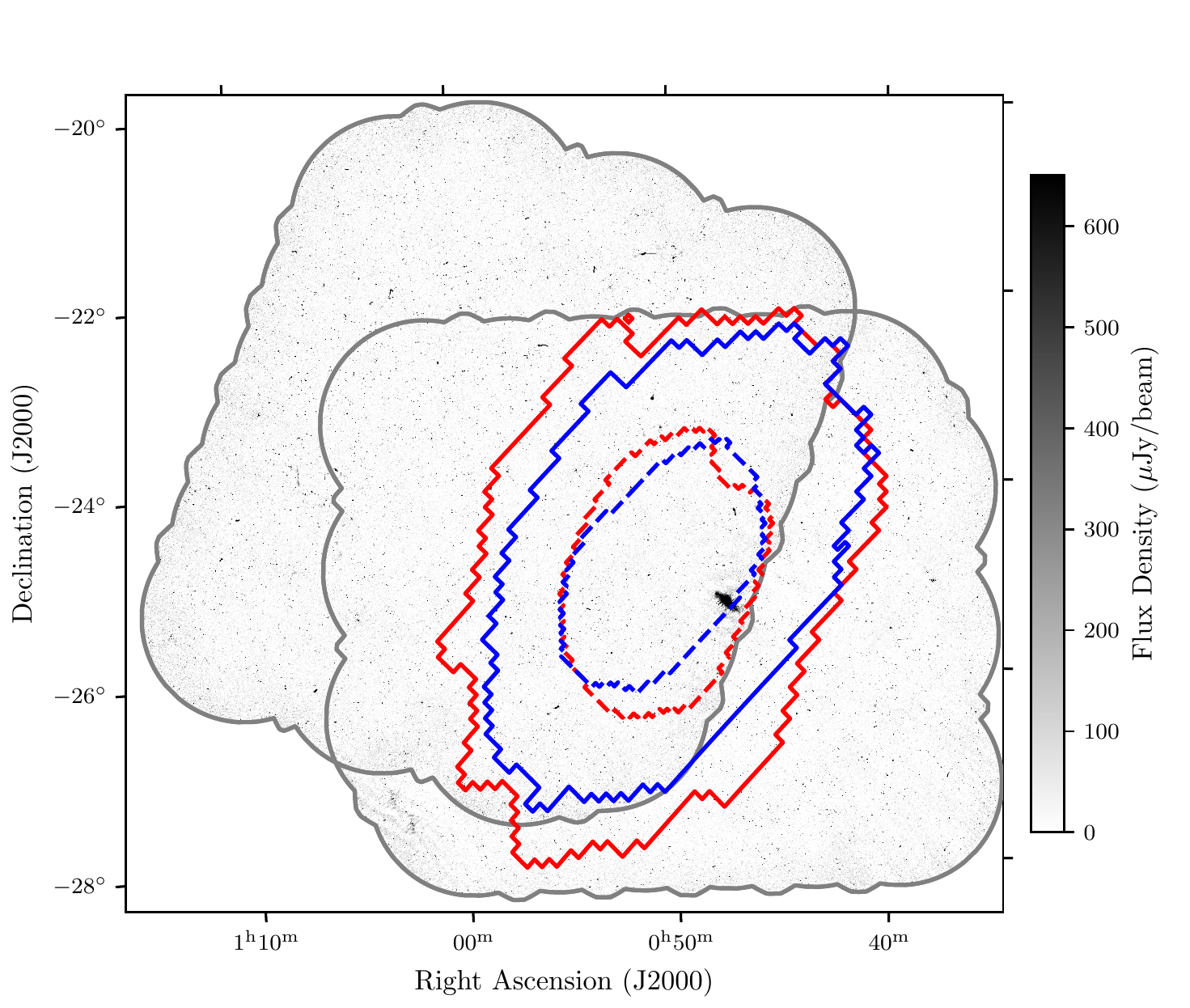}
    \caption{ASKAP imaging of the localisation region of GW190814 at 234 (bottom right) and 260 (top left, rotated as discussed in Section \ref{sec:data}) days post-merger with both footprints outlined in grey. The 50\% (dashed) and 90\% (solid) contours for the initial and final skymaps are shown in red and blue respectively.}
    \label{fig:localisation}
\end{figure*}

\section{Data and Analysis}
\label{sec:data}
\subsection{Observations and Data Reduction}
Table \ref{tab:obs_descrip} provides the details of our ASKAP observations, which were carried out at 944\,MHz and span 2 to 655 days post-merger. We used the closepack36\footnote{See \citet{2021PASA...38....9H}} footprint with a beam spacing of 0.9$\degr$ for all observations, centered on $\alpha=00^{\rm h}50^{\rm m}37\fs5$, $\delta=-25\degr16\arcmin57\fs37$ (J2000) corresponding to the posterior maximum of the initial skymap as seen in Figure \ref{fig:localisation}. This footprint covers 89\% and 87\% of the initial and final gravitational wave localisations respectively, with large extraneous coverage. Each observation was approximately 10 hours, achieving full u-v coverage along with a typical sensitivity of 35--40\,$\mu$Jy. However, there are three exceptions to the above specifications:

\begin{enumerate}
    \item Epoch 5 consists of two observations separated by 1 day, as a result of technical difficulties encountered five hours into the initial observation. We have combined the good data from each observation into a single image consisting of $\sim 15$ hours on-source, resulting in better sensitivity.
    \item Epoch 6 was centered on $\alpha=00^{\rm h}58^{\rm m}00$, $\delta=-23\degr45\arcmin00$  (offset by $\sim 2\degr$) and rotated with respect to the other pointings by 67.5$\degr$ in order to rule out instrumental effects as the origin of six rapid scintillators discovered in the field \citep{2021MNRAS.502.3294W}. The footprint of this pointing has a 74\% overlap with the primary footprint and also covers 89/87\% of the initial/final localisation\footnote{Both pointings have similar total coverage due to the large extraneous area covered by each. The shifted pointing covers the entirety of the 50\% credible interval for this lobe, and the majority of the 90\% credible interval.}.
    \item Epoch 9 consists of two $\sim 3.5$ hour observations with a 3.5 hour gap between them. We have combined the data from each observation into a single image consisting of $\sim 7$ hours on source.
\end{enumerate}

Each observation was reduced using the ASKAPsoft pipeline \citep{2017ASPC..512..431W} with standard parameters, as described by \citet[][]{2019ApJ...887L..13D}.
To assess the data quality we selected all bright (SNR>7), isolated (no sources within 
150\,arcsec) compact sources, where we following the definition of compactness by
\citet{2021arXiv210900956H} of an integrated to peak flux ratio of $S_I/S_P < 1.024 + 0.69\times \textrm{SNR}^{-0.62}$,
and compared their position and flux density in the first epoch to all subsequent observations. We find a median peak flux density ratio of $0.99 \pm 0.11$ and median positional offsets of $0.01\pm 0.70$ and $-0.03\pm 0.61$ arcsec in Right Ascension and Declination respectively.

\subsection{Transient Search}
We have used all 10 epochs to undertake a search for all intrinsically transient radio sources in the field. These observations are the most sensitive radio imaging of this field to-date, so no pre-merger reference image exists. We therefore cannot exclude the possibility that the radio counterpart to GW190814 was detected in our first observation and as such we, do not require a constraining non-detection of a source for it to be considered a candidate counterpart, or have a transient origin. This definition also ensures that our search is sensitive to all transient sources in the field, independent of our observing strategy. However, it does mean that our search will initially be contaminated with variable sources (either intrinsic variables like pulsars and AGN, or steady sources that are varying due to scintilation). In addition, we distinguish between intrinsic transients (those originating from a one-off, cataclysmic event) and observational transients (variable sources that appear transient due to our observing sensitivity and cadence). We refer to the former simply as ``transients'' throughout the rest of the text.

\subsubsection{Untargeted radio variability search}
\label{subsec:eta-v-search}
We carried out a transient search of the field using the VAST transient detection pipeline\footnote{\url{https://vast-survey.org/vast-pipeline/}} \citep{2021arXiv210806039M,2021arXiv210105898P} using the standard ASKAPSoft data products. We used a de Ruiter radius of 5.68 \citep[][]{2011PhDT.......427S} to associate sources between epochs and recalculated the uncertainty estimates produced by the ASKAPSoft source finder, {\sc selavy}, using the equations of \citet{1997PASP..109..166C}.

We applied the following criteria to build our variability search source sample:
\begin{itemize}
    \item A ratio of integrated to peak flux density $<1.5$
    \item Maximum signal-to-noise ratio in a single epoch larger than 5
    \item 2 selavy detections
    \item No relations\footnote{See \url{https://vast-survey.org/vast-pipeline/design/association/##relations}}
    \item Distance to nearest source $>1\,$\arcmin
\end{itemize}
which reduced our sample size from 66\,117 to 10\,254. We note that the final two criteria mean that this search is not sensitive to mergers ocurring in radio-loud hosts. While this scenario is unlikely \citep{2016ApJ...831..190H}, for completeness we have carried out a galaxy-targeted search independent of the above criteria which we describe in Section \ref{subsec:galaxy-targeting}.

\begin{figure}
    \centering
    \includegraphics{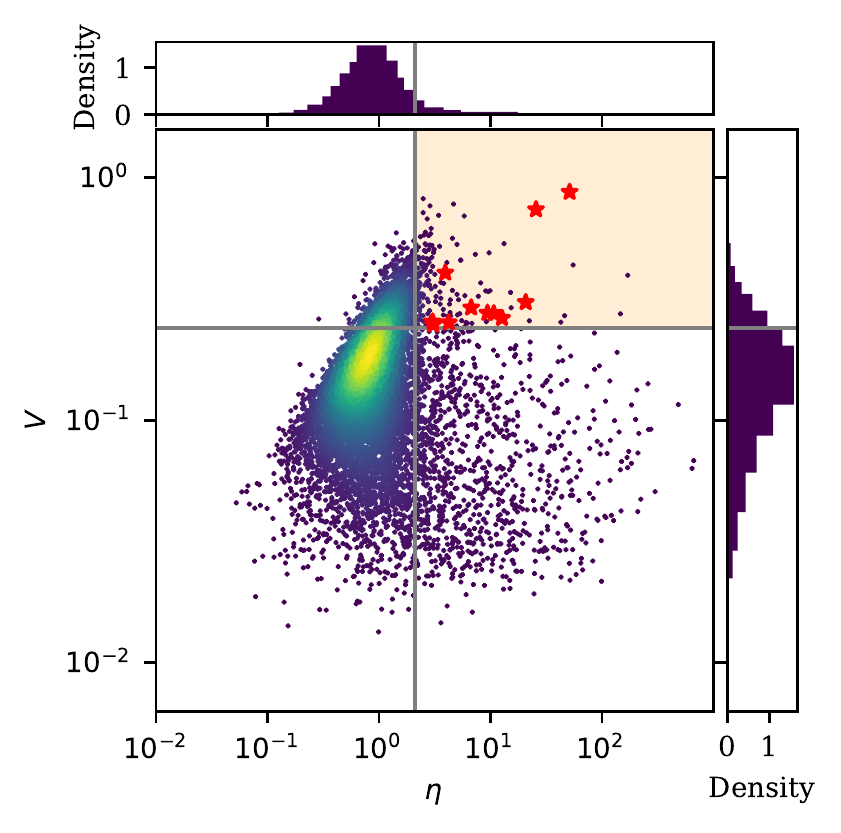}
    \caption{Variability metrics ($V$ and $\eta$) for all sources in our sample, coloured by kernel density estimate. Grey lines show the cutoffs used for each metric, while the shaded quadrant (top right) shows the sources that are considered to be statistically variable. These sources were manually vetted to search for transient candidates, which are denoted by red stars.}
    \label{fig:eta-v}
\end{figure}

We calculated the standard $\eta$ and $V$ variability metrics for the peak flux density of the remaining sources in our sample, defined as

\begin{equation}\label{e_var2}
V = \frac{1}{\overline{S}}\sqrt{\frac{N}{N-1} (\overline{S^2} - \overline{S}^2)},    
\end{equation}
\begin{equation}\label{e_var1}
\eta = \frac{N}{N-1}\left( \overline{w S^2} - \frac{\overline{w S}^2}{\overline{w}}\right),
\end{equation}
where $N$ is the number of observations, $S_i$ and $\sigma_i$ are the flux density and uncertainty in the $i$th epoch and overbars denote means (i.e., $\overline{S} \equiv \frac{1}{N}\sum_i S_i$). We then fit a Gaussian to the distribution of the logarithm of each metric to calculate the mean, $\mu$, and standard deviation, $\sigma$ (see Figure \ref{fig:eta-v}). Based on \citet{2019A&C....27..111R} we consider sources with $\eta > \mu_\eta + 1.5\sigma_\eta$ and $V > \mu_V + 1.0\sigma_{V}$ to be significantly variable, corresponding to $\eta > 2.1$ and $V > 0.24$ for this dataset. This resulted in 186 candidate variable sources.

We then manually classified all sources via inspection of their lightcurves and images. We found that 81 candidates were clearly artefacts. These sources are either sidelobes of bright sources, noise spikes incorrectly classified as sources, or sources that are at the near the edge of the epoch 6 footprint and do not pass the variability threshold after removing that measurement. We classified an additional 84 candidates as either potential artefacts or variable sources detected at low significance. Finally, we found that 10 candidates with lightcurves that are consistent with persistent radio sources exhibiting variability. Of these, one is a known pulsar (PSR J0038$-$2501), while the remaining nine have infrared counterparts in the Wide-field Infrared Survey explorer All-sky data release \citep[WISE;][]{2012yCat.2311....0C}. This leaves 11 sources with lightcurves that exhibit a rise and fall consistent with expectations for a radio transient. We discuss these in detail in Section \ref{subsec:candidate_analysis}

\subsubsection{Galaxy-targeted search}
\label{subsec:galaxy-targeting}
We also carried out an independent galaxy-targeted search using version 2.4 of the GLADE catalogue \citep[][]{2018MNRAS.479.2374D}. We searched for GLADE sources within 20\arcsec of all 66\,117 sources found by the VAST pipeline. This crossmatch radius corresponds to a physical distance of 23\,kpc at the estimated distance to the merger (241\,Mpc), larger than the host galaxy offset of >90\% of known short GRBs \citep{2013ApJ...776...18F}. We then removed all GLADE sources with distance estimates that are outside of the 90\% credible interval of the distance to the merger, leaving 662 unique galaxies, of which 325 have no distance estimate. This corresponds to a sample of 799 VAST pipeline sources. Since we are using Stokes I total intensity images, any real source must have a positive flux density, and therefore must have a positive variability index. After applying this constraint we were left with 589 sources.

We manually inspected the lightcurves and images of all 589 sources with the goal of searching for counterparts that may have been missed by the search described in Section \ref{subsec:eta-v-search} due to having small offsets from a radio-loud host galaxy. We found two sources that are not artefacts and have lightcurves that are qualitatively consistent with the rise and fall expected of an extragalactic radio transient, one of which (ASKAP~J003537.3$-$271844) was already discovered in the search described in Section \ref{subsec:eta-v-search}.

\section{Discussion}
\label{sec:discussion}

\begin{figure*}
    \centering
    \includegraphics{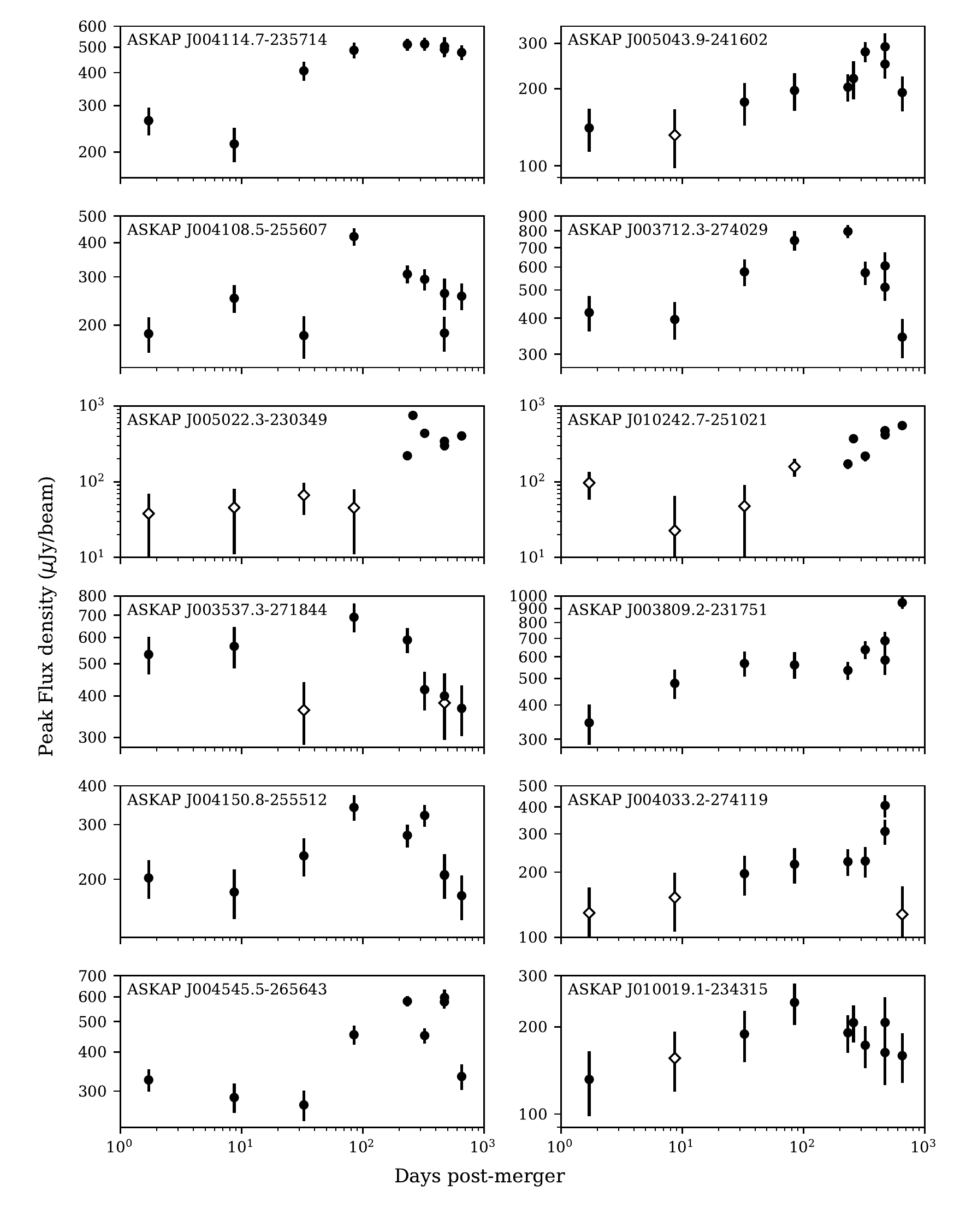}
    \caption{Lightcurves for the 12 candidate afterglows found in our initial search. Source-finder measurements are denoted by solid circles, while open diamond markers show forced photometry on images with no detection.}
    \label{fig:full_lightcurves}
\end{figure*}

\subsection{Candidates}
\label{subsec:candidate_analysis}
Figure \ref{fig:full_lightcurves} shows the lightcurves of the 12 candidate sources found in our search (11 from our untargeted search, plus one additional source from the galaxy-targeted search). We have used a number of archival surveys to help classify each candidate. The first epoch of the Rapid ASKAP Continuum Survey \citep[RACS;][]{2020PASA...37...48M} was at an observing frequency of 888\,MHz and covers the sky south of $+41\,\degr$ to a sensitivity of $\sim 250\,\mu$Jy. The fields of interest for our work were observed on 2019-04-27 and 2019-04-28 (before the gravitational wave event) and there are no crossmatches with our candidates. The VLA Sky Survey \citep[VLASS;][]{2020PASP..132c5001L} is an ongoing survey at 3\,GHz covering the sky north of $-40\,\degr$ to a sensitivity of 120\,$\mu$Jy in each epoch. The fields of interest for our work were observed in epochs 1.1, 1.2 and 2.1 on 2018-02-18, 2019-07-09, and 2020-10-27 respectively (with the first two observed prior to GW190814). Parts of this field were also covered by the GW190814 follow-up reported by \citet{2021arXiv210208957A} and one candidate (ASKAP~J005022.3$-$230349) is within the primary beam of any pointing.

To compare the variability of the source to the expected extrinsic variability caused by scintillation we use NE2001 \citep[][]{2002astro.ph..7156C} to calculate the Galactic contribution to the electron distribution along the line of sight and the equations of \citet[][]{1998MNRAS.294..307W} to calculate the expected variability due to scintillation. We find that compact sources in this field at our observing frequency will exhibit $\sim 30\%$ variability on timescales of $\sim 10$ days.

\begin{figure}
    \centering
    \includegraphics{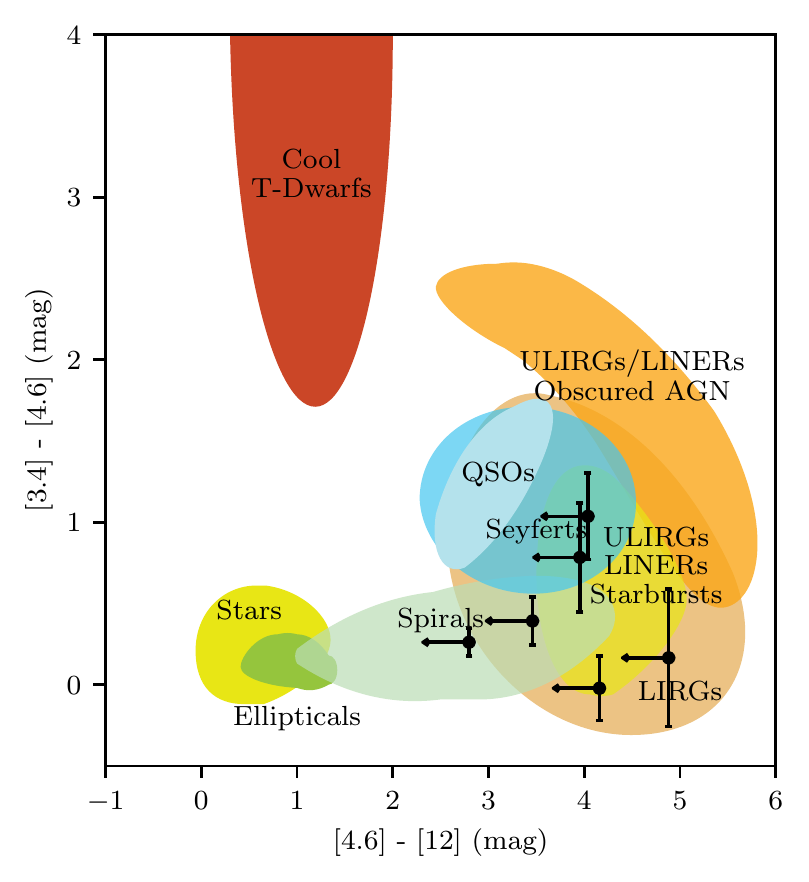}
    \caption{Colour-colour diagram of the WISE sources associated with the candidate variables and transients discussed in Section \ref{subsec:candidate_analysis}. We have also included the classification regions from \citep{2010AJ....140.1868W} for context, although we note that any classification of these sources from WISE colours alone is limited by a lack of any detection in the 12$\mu$m band.
}
    \label{fig:wise_colourcolour}
\end{figure}

We use Data Release 8 of the Dark Energy Spectroscopic Instrument Legacy Imaging Surveys \citep[][]{2019AJ....157..168D}, specifically data from the Dark Energy Camera Legacy Survey (DECaLS), to search for optical emission associated with our candidates or their potential hosts, along with the photometric redshift estimates from \citet[][]{2021MNRAS.501.3309Z}. We have also used the WISE All-sky data release to search for infrared emission associated with our candidates and calculated the standard WISE colours, shown in Figure \ref{fig:wise_colourcolour}. However, no sources have a detection in the WISE 12\,$\mu$m band and therefore any decisive classification from colours alone is not possible.

\subsubsection{ASKAP~J004114.7$-$235714}
ASKAP~J004114.7$-$235714 is within the 90\% credible interval for the localisation of GW190814 and doubles in flux density between the first and fourth epochs before remaining relatively stable in the remaining observations. The coordinates are covered by VLASS 1.2, but no radio emission is detected. The source has a counterpart in both WISE and DECaLS, with a photometric redshift of $1.0\pm 0.1$. The source has a variability index of $V=0.26$, consistent with expectations for refractive scintillation along this line of sight. We therefore suggest that this source is an unrelated variable and rule it out as a counterpart.

\subsubsection{ASKAP~J005043.9$-$241602}
ASKAP~J005043.9$-$241602 is within the 30\% credible interval for the localisation of GW190814. The source was observed by VLASS 1.2, but no radio emission is detected. The closest DECaLS source is offset by $14\arcsec$, with a photometric redshift of $z=1.4\pm 0.5$ (corresponding to a physical offset of $\sim 120$\,kpc) and is therefore unrelated to the radio source. While we cannot classify this source using existing archival data, it is unlikely to be a transient due to the low variability index ($V=0.26$) that is consistent with refractive scintillation. However, we cannot comprehensively rule it out.

\subsubsection{ASKAP~J004108.5$-$255607}
ASKAP~J004108.5$-$255607 is outside of the 99\% credible interval for the localisation of GW190814 and we therefore rule it out as a counterpart. It is spatially coincident with sources in WISE and DECaLS, with a photometric redshift of $z=0.88\pm 0.05$. The source was observed in VLASS 1.2, but no radio emission was detected. The source has a variability index of $V=0.29$, consistent with expectations for refractive scintillation along this line of sight. We therefore suggest that this source is likely an unrelated variable.

\subsubsection{ASKAP~J003712.3$-$274029}
ASKAP~J003712.3$-$274029 is outside of the 99\% credible interval for the localisation of GW190814 and we therefore rule it out as a counterpart. It is spatially coincident with sources in WISE and DECaLS, with a photometric redshift of $1.7\pm 0.5$. The source has a variability index of $V=0.28$, consistent with expectations for refractive scintillation along this line of sight. There is a flux excess of $\sim 800\,\mu$Jy at the source location in VLASS 1.2, implying that the source was significantly brighter prior to our first epoch, unless it has an optically thin spectrum. We therefore comprehensively rule this source out as a transient and classify it as an unrelated variable.

\subsubsection{ASKAP~J005022.3$-$230349}
\label{subsec:candidate}
ASKAP~J005022.3$-$230349 tripled in luminosity between epochs 5 and 6 before slowly declining. While the final epoch shows a slight rise compared to the trend of the previous data points, this deviation from the trend is consistent with the expected variability due to scintillation along this line of sight.

The candidate position is covered by the publicly available VLA observations reported by \citet{2021arXiv210208957A} on 2019-09-22 and 2020-02-29, which have independently analysed. Data were calibrated using the automated pipeline available in the Common Astronomy Software Applications (CASA; \citealt{McMullin2007}), with additional flagging performed manually, and imaged\footnote{Cell size was 1/10 of the synthesized beamwidth, field size was the smallest magic number ($10 \times 2^n$) larger than the number of cells needed to cover the primary beam.} using the CLEAN algorithm \citep{Hogbom1974}. Because the source is offset from the phase centre we also performed a primary beam correction with {\sc pbcorr}. We find no radio emission to  a $3\sigma$ limit of 36$\,\mu$Jy at 6\,GHz in either epoch.

The candidate is offset by 1.9\arcsec\ from an optical source that has been observed by DECaLS and Gaia \citep[][]{2016A&A...595A...1G}. The catalogued photometric redshift based on DECaLS data is $z=0.3$ \citep[95\% confidence interval 0.09-0.7;][]{2021MNRAS.501.3309Z}, which is consistent with the redshift of the merger, and the source has no significant parallax measurement in the second Gaia Data Release \citep{2018A&A...616A...1G}. We obtained spectroscopy of the optical source on 2020-12-08 with the Robert Stobie Spectrograph \citep[RSS;][]{burgh03} on the 10m-class Southern African Large Telescope (SALT).  We find no evidence for any host galaxy emission lines. Instead, the spectrum is consistent with that of an M-dwarf star. We have since queried the Gaia early DR3 \citep[released after our SALT observations had been carried out;][]{2021A&A...649A...1G} and find a parallax consistent with a distance of $\sim$400\,pc. We therefore rule out the optical source as the host galaxy of the candidate. We also note that the M dwarf is not the source of the radio emission based on the the spatial offset and the variability timescale \citep[radio flare stars are generally variable on timescales of minutes--hours, see e.g.][]{2020ApJ...905...23Z}.

\begin{figure}
    \centering
    \includegraphics[width=\linewidth]{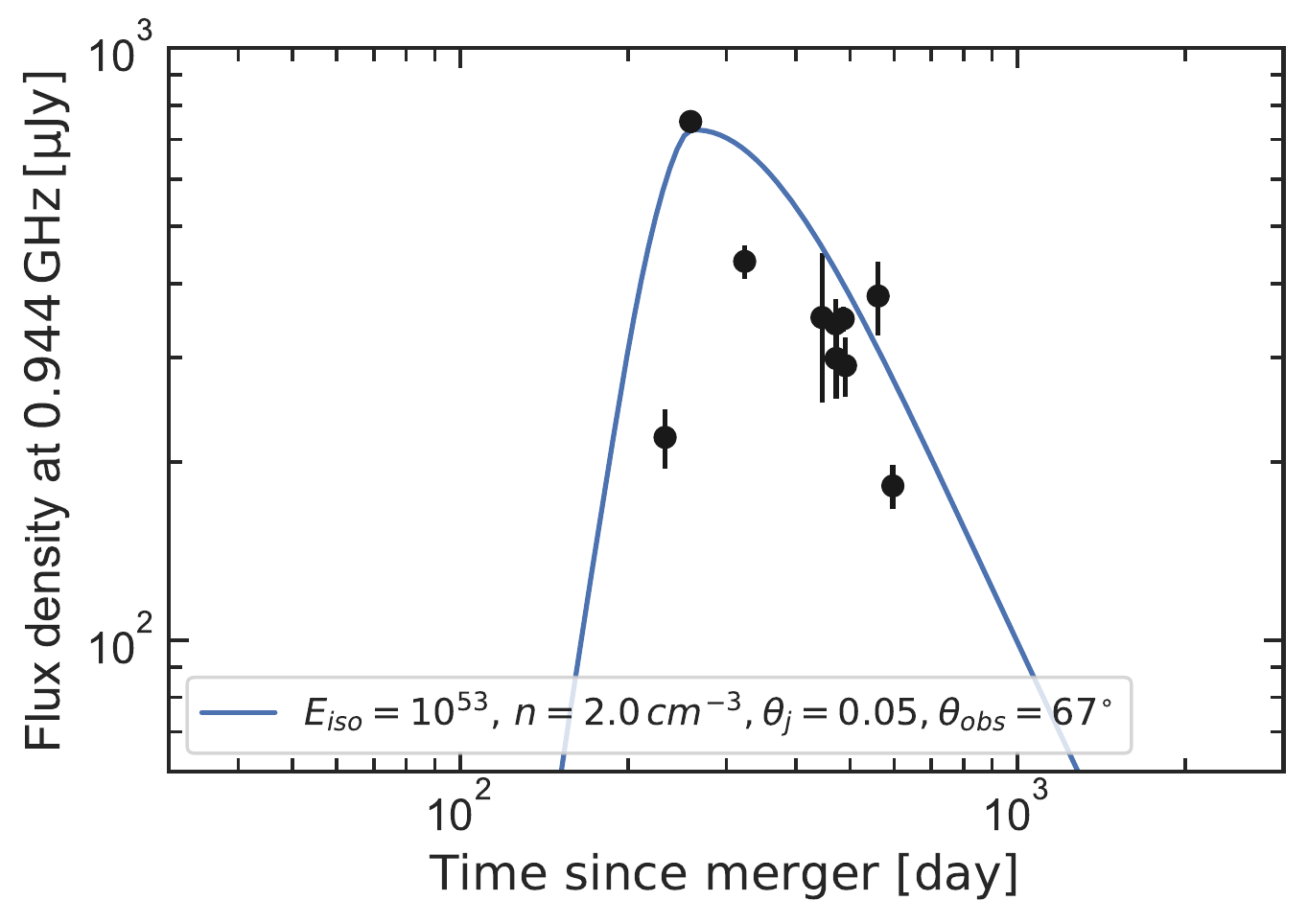}
    \caption{Detctions of ASKAP~J005022-230348 in our GW190814 follow-up observations. The lightcurve of a tophat jet with with an isotropic equivalent energy of $E_{\rm iso}=10^{53}$\,erg propagating into a medium with density $n=2\,$cm$^{-3}$ viewed at an angle of 67\,\degr\ off-axis is shown in blue. Even this extreme assumption cannot reproduce the rapid rise observed for this source.}
    \label{fig:190814_candidate_lc}
\end{figure}

We find that even the most extreme assumption of a highly energetic ($E_{\rm iso}=10^{53}$\,erg) top-hat jet (i.e. a jet with uniform velocity profile, as opposed to structured jets) propagating into a dense ($n=2\,$cm$^{-3}$) circum-merger medium cannot reproduce the observed steep rise (see Figure \ref{fig:190814_candidate_lc}). We conclude that this source is not a viable counterpart to GW190814, but do consider it to be a real (but unrelated) radio transient that likely occurred months after the merger. Detailed study and classification of this source is beyond the scope of this work and will be presented in a separate manuscript (Dobie et al. in prep.).

\subsubsection{ASKAP~J010242.7$-$251021}
ASKAP~J010242.7$-$251021 is outside of the 99\% credible interval for the localisation of GW190814 and we therefore rule it out as a counterpart. It is spatially consistent with a DECaLS source with photometric redshift $z=0.3\pm 0.1$. The source has constraining non-detections in our first four epochs, although manual inspection of the images suggests marginal evidence for a detection in the fourth epoch. The source has a variability metrics of $V=0.74$ and $\eta = 25$. However, it is detected in VLASS 1.2 with a flux density of $\sim 700\,\mu$Jy, 270 days before it is detected in our observations. Extrapolating the observed trend in the 944\,MHz lightcurve, we find that the source would have been a few $\mu$Jy at the time of the VLASS observation, corresponding to a spectral index of $\alpha\gtrsim 5$.  We therefore rule this out as a transient  and instead classify it as an intrinsically variable source.

\subsubsection{ASKAP~J003537.3$-$271844}
ASKAP~J003537.3$-$271844 was found in both our untargeted variability search and our galaxy targeted search. However, it is outside the 99\% credible interval for the localisation GW190814 and we therefore rule it out as a counterpart. It is offset by 1.1\arcsec\ from a known galaxy with a spectroscopic redshift of $z=0.22436$ observed as part of the 2-degree Field Lensing Survey \citep[][]{2016MNRAS.462.4240B}. This galaxy also has an infrared counterpart detected by WISE. There is marginal evidence for emission at the radio source location in VLASS at $\sim 480\,\mu$Jy although we caution that this is only three times the local noise level and there are multiple pixels within a 30\arcsec\ radius with comparable flux density measurements. While the positions of the radio source and the galaxy are marginally discrepant ($\gtrsim 1\sigma$), the inferred radio luminosity of the radio source if it is at the distance of the galaxy is $\sim 10^{30}$\,erg\,s$^{-1}$\,Hz$^{-1}$, consistent with the population of AGN \citep{2011ApJ...731L..41B}. We therefore conclude that this source is likely an AGN.

\subsubsection{ASKAP~J003809.2$-$231751}
ASKAP~J003809.2$-$231751 is outside the 98\% credible region for the localisation of GW190814 and is therefore unlikely to be a counterpart. There is no evidence for radio emission in either VLASS 1.1 or 2.1. The radio lightcurve consists of a continual rise, with the flux density almost tripling across 10 epochs and a variability index of $V=0.27$. The source is offset from two sources in DECaLS with separations of 2.9\arcsec\ and 3.8\arcsec\ and photometric redshifts of $z=0.2\pm 0.1$ and $z=0.8\pm 0.2$. These angular separations at those redshifts correspond to physical offsets of 10\arcsec\ and 30\arcsec\ respectively, meaning that either optical source could plausibly be the host galaxy if the radio source is a transient.

However, we note that the source exhibits no significant variability between epochs 2 and 9. Applying the two-epoch variability metrics of \citet{2016ApJ...818..105M}, only epochs 1 and 10 show significant variability when compared to other epochs. Removing either epoch from our analysis results in a variability metric, $V$, lower than the cutoff in Section \ref{subsec:eta-v-search}. Based on this and the fact that the variability metric of the full radio lightcurve is comparable to expectations for refractive scintillation, we suggest that it is unlikely that this source is a transient. However, we are unable to classify it with archival data, nor comprehensively rule it out.

\subsubsection{ASKAP~J004150.8$-$255512} 
ASKAP~J004150.8$-$255512 is outside of the 99\% credible interval for the localisation of GW190814 and we therefore rule it out as a counterpart.  It is spatially coincident with a source in both WISE and DECaLS, with photometric redshift $z=1.1\pm0.1$. There is no counterpart in epoch 1.2 of VLASS, observed on 2019-07-06. The variability index of this source is $V=0.25$, consistent with expectations for refractive scintillation. We therefore suggest that this source is likely an unrelated variable.

\subsubsection{ASKAP~J004033.2$-$274119} 
ASKAP~J004033.2$-$274119 is outside of the 99\% credible interval for the localisation of GW190814 and we therefore rule it out as a counterpart. It is spatially coincident with sources in WISE and DECaLS, with photometric redshift $z=0.7\pm0.05$. There is marginal ($\sim 3\sigma$) evidence for radio emission of $\sim 400\,\mu$Jy in VLASS 1.2, observed on 2019-07-06. While the variability index ($V=0.4$) of this source is higher than expectations for refractive scintillation, the spatial coincidence with a known optical and IR source suggests that it is likely an AGN exhibiting some combination of intrinsic and extrinsic variability.

\subsubsection{ASKAP~J004545.5$-$265643}
ASKAP~J004545.5$-$265643 is outside of the 99\% credible interval for the localisation of GW190814 and we therefore rule it out as a counterpart. There is no coincident optical source in DECaLS, and the closest has a photometric redshift of $1.7\pm0.7$. However, the optical source is offset by 12\arcsec\ and is therefore unlikely to be related to the radio source. The source has a likely counterpart in VLASS 1.2 with flux density $\sim 500\,\mu$Jy suggesting that it was brighter prior to our observations and is therefore not a transient. The variability index of $V=0.3$ is consistent with expectations for a compact source exihbiting refractive scintilation.

\subsubsection{ASKAP~J010019.1$-$234315}
ASKAP~J010019.1$-$234315 was found in our galaxy targeting search and is outside of the 99\% credible interval for the localisation of GW190814 so we rule it out as a counterpart. It is offset from PGC\,3195680, which has no distance estimate in GLADE, by 17\arcsec. After querying WISE and DECaLS we find that there is a closer source offset by only 1.3\arcsec\ with a photometric redshift of $z=1.2\pm 0.25$. The source is covered by VLASS 1.1 and 2.1 but no radio emission is detected in either epoch, which is not unexpected given its low flux density. Similar to ASKAP~J003537.3$-$271844, the offset between the radio and optical sources is small, but statistically significant and the inferred radio luminosity ($\sim 10^{30}$\,erg\,s$^{-1}$\,Hz$^{-1}$) is broadly consistent with expectations for an AGN. We therefore suggest that this source is also an AGN, but note that the classification is less certain in this instance due to the larger spatial offset, alongside the larger uncertainty of the photometric redshift.

\begin{figure}
    \centering
    \includegraphics{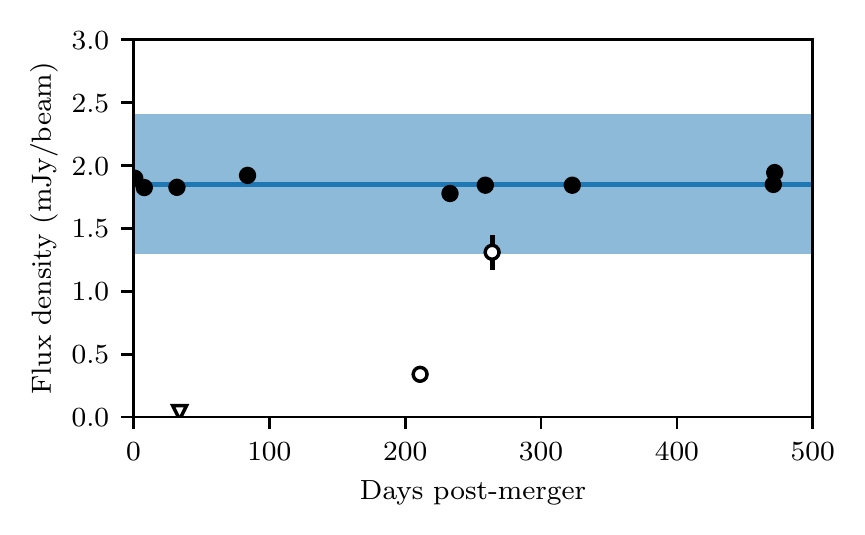}
    \caption{Radio lightcurve of the candidate near ESO~474-035 reported by \citet{2021arXiv210208957A}. ASKAP observations at 944\,MHz are denoted by solid markers, while open markers show VLA observations at 6\,GHz and triangle correspond to 3$\sigma$ upper limits. The blue line shows the mean flux density in our observations, while the maximum expected variability due to scintillation at 944\,MHz ($\sim$30\%) is denoted by the blue shaded region.}
    \label{fig:alexander_candidate_lc}
\end{figure}

\subsection{Candidate counterpart associated with ESO 474-035}
\citet{2021arXiv210208957A} reported the discovery of a candidate radio counterpart near ESO~474-035 in their galaxy-targeted follow-up observations. The observed lightcurve and spectral energy distribution is consistent with a highly energetic ($E_{\rm iso}\sim 8\times 10^{53}$) tophat jet propagating through a dense medium ($n\sim 0.5\,$cm$^{-3}$), or high velocity ($\beta_0\sim 0.8c$) kilonova ejecta. The authors suggest that the candidate is unrelated to GW190814 as the required energies and velocities are high compared to the population of known short GRBs and compact object mergers.

The source is detected in all nine ASKAP observations with an average peak flux density of 1.85\,mJy, in good agreement with the spectral energy distribution reported by \citet{2021arXiv210208957A}. The lightcurve (Fig. \ref{fig:alexander_candidate_lc}) is consistent with a relatively steady source and we measured $V_s=0.03$, $\eta=3.04$. We can therefore comprehensively rule it out as a counterpart to GW190814, independent of any physical arguments.

\begin{figure}
    \centering
    \includegraphics[width=\linewidth]{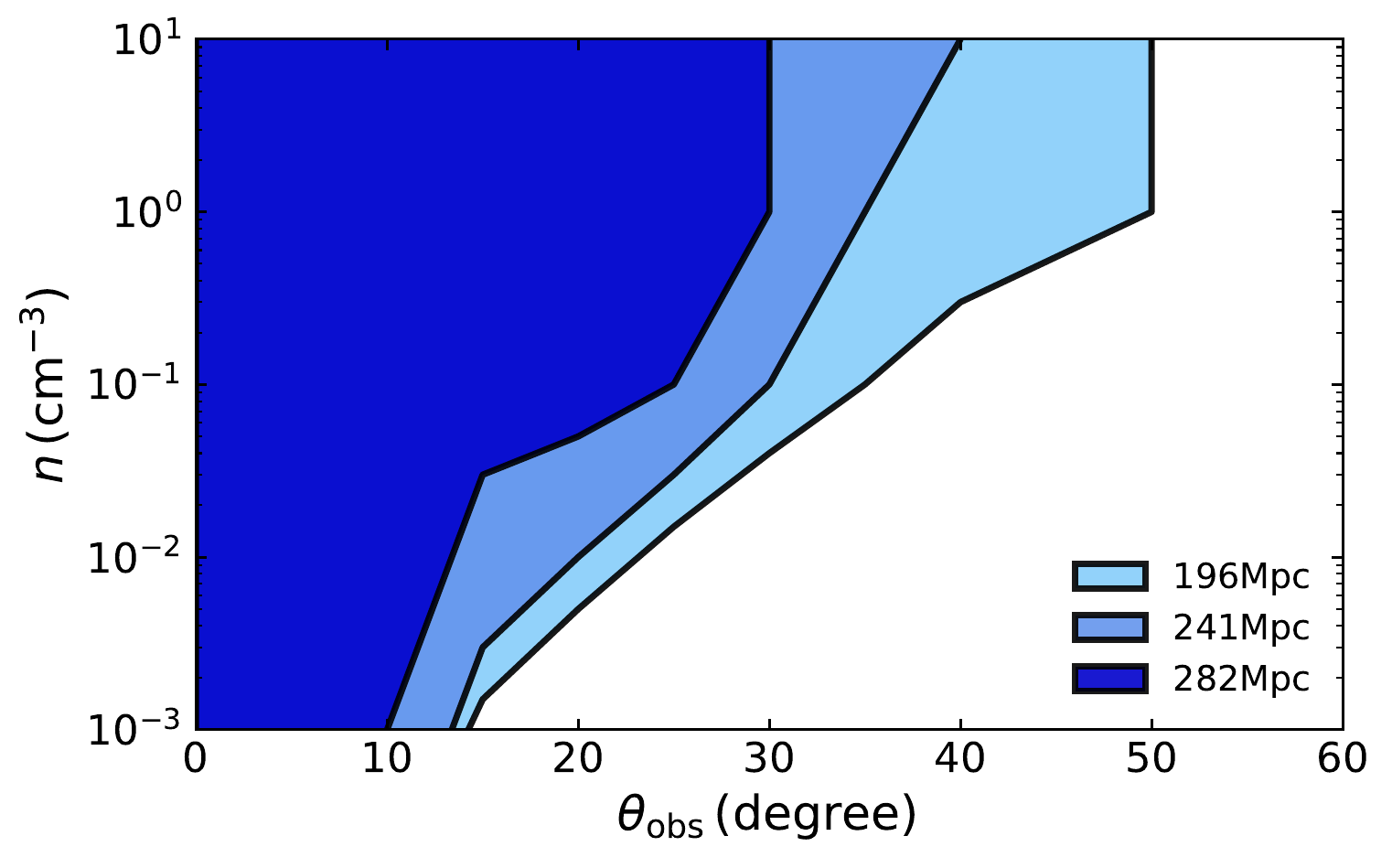}
    \caption{Parameter space ruled out by our radio non-detections for a merger with isotropic equivalent energy $10^{51}$\,erg, an initial jet opening angle of $10\degr$ and microphysics parameters $\epsilon_{e}=0.1$, $\epsilon_{B}=0.01$ and $p=2.2$. Shaded regions correspond to the ruled out parameter space for a range of distances corresponding to $1\sigma$ either side of the median as determined by gravitational wave measurements.}
    \label{fig:kenta_constraints}
\end{figure}

\subsection{Constraints on the properties of a relativistic jet associated with GW190814}
Based on a continued non-detection of a radio afterglow from GW190814 we can constrain the physical properties of any potential outflow from the merger. We do this using two approaches. 

In \citet{2019ApJ...887L..13D} we constrained the inclination angle and circum-merger density of the merger using afterglow lightcurves from an off-axis tophat jet with isotropic equivalent energy $E_{\rm iso}=10^{51}$\,erg, a jet opening angle of $10\degr$ and microphysics parameters $\epsilon_{e}=0.1$, $\epsilon_{B}=0.01$ and $p=2.2$. Figure \ref{fig:kenta_constraints} shows the same procedure applied to our more recent results along with the updated gravitational wave distance estimate. We find a significant improvement over our previous results, and are able to rule out an additional $\sim$10\,\degr\ of parameter space across all values of circum-merger density. Comparing our results to the merger inclination angle estimated from the gravitational wave signal \citep[45\,\degr;][]{2020ApJ...896L..44A} we find that our results are only constraining for the lower end of the distance estimate, where we are able to constrain $n\lesssim 0.5$\,cm$^{-3}$.

\begin{figure*}
    \centering
    \includegraphics{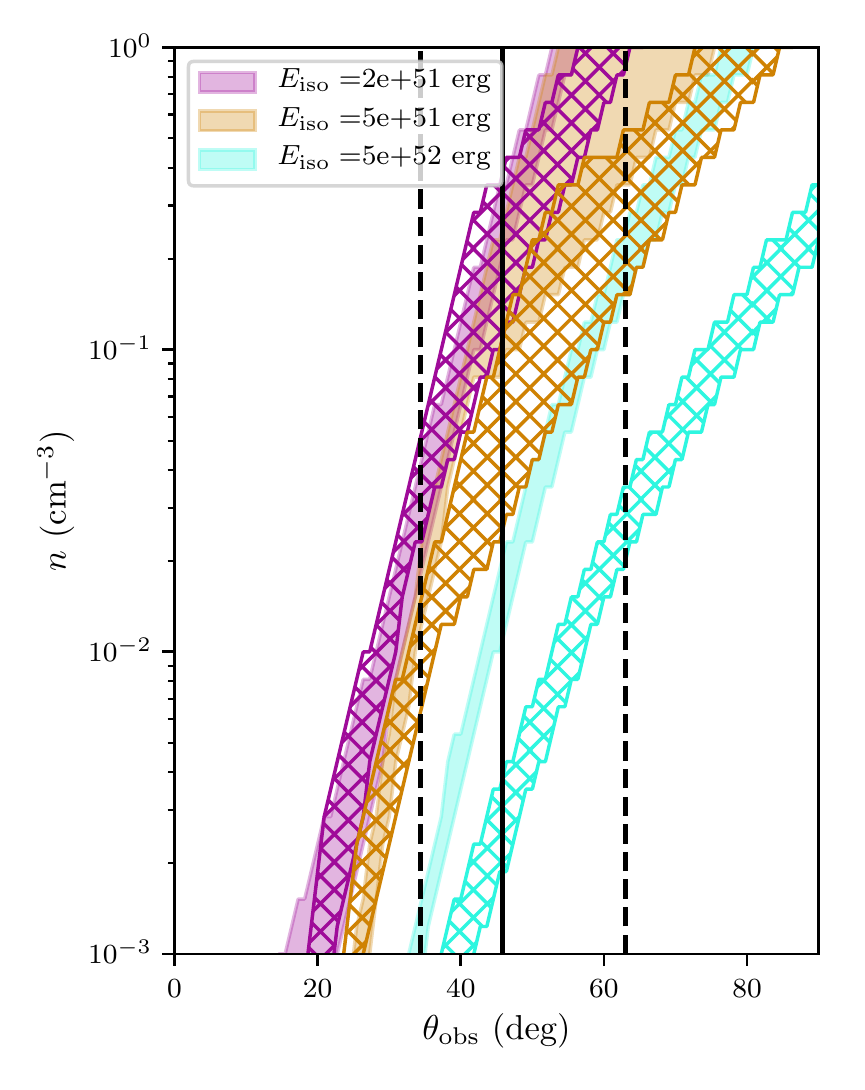}
    \includegraphics{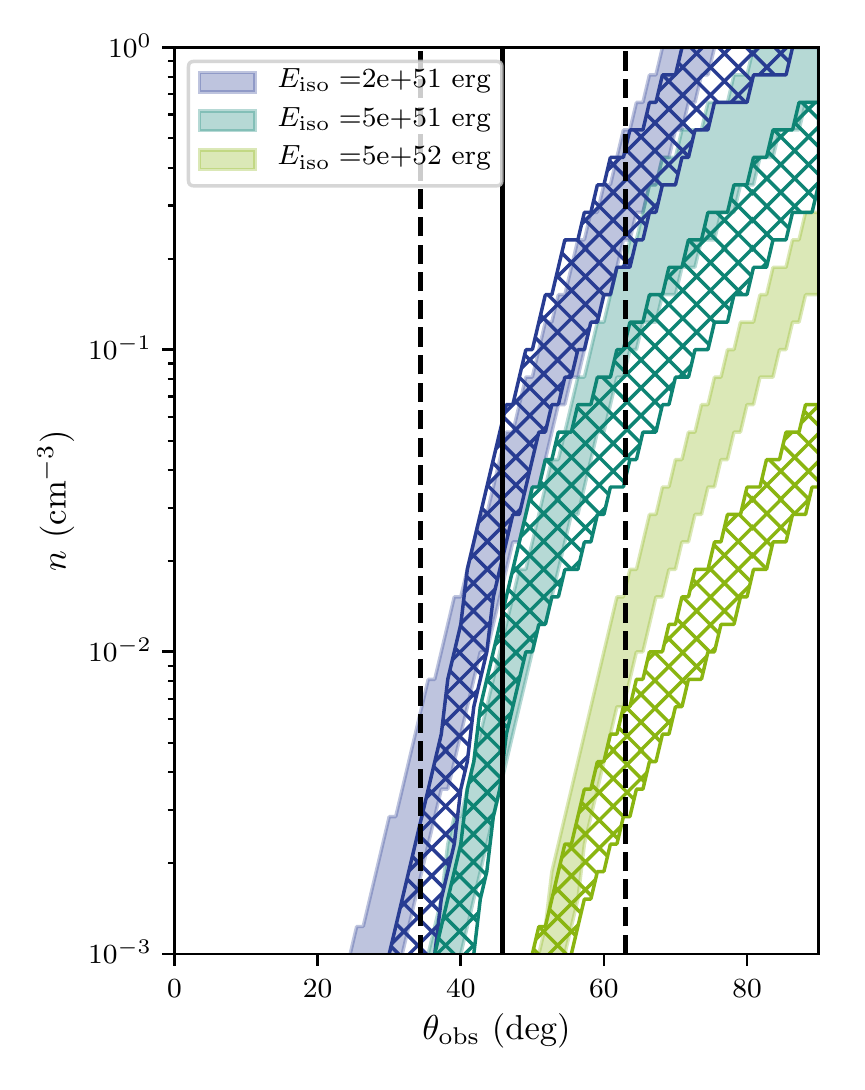}
    \caption{Parameter space ruled out by our radio non-detections to date for a 15$\degr$ tophat (left) and Gaussian (right) jet. Shaded regions show constraints from \citet{2021arXiv210208957A}, while hatched regions show the constraints from this work, with the areas to the upper left of the region ruled out. The width of the region corresponds to the uncertainty in the distance to the merger. The inclination angle and associated $1\sigma$ uncertainties from the gravitational wave signal are shown in the solid and dashed vertical lines respectively.}
    \label{fig:alexander_comparison}
\end{figure*}

\citet{2021arXiv210208957A} follow a similar logic using {\sc afterglowpy} \citep[][]{2020ApJ...896..166R} for tophat and Gaussian jet models and two 6\,GHz non-detections at 38 and 208 days post-merger. Both models assume the same microphysics parameters as above with a jet opening angle of $15\degr$ and are computed for a range of isotropic equivalent energies ($E_{\rm iso}=2\times 10^{51}$, $5\times 10^{51}$, $5\times 10^{52}$\,erg). The Gaussian jet model consists of a core with wings extending to $90\degr$. We have applied the same process to our observations and Figure \ref{fig:alexander_comparison} shows our limits which are comparable to \citet{2021arXiv210208957A}, albeit with slightly better constraints for higher energies. We note that our limits are more comprehensive due to the substantially higher fraction of the localisation area covered by our observations.

\subsection{Radio Transient Rates}
\label{subsec:rtr}
In \citet[][]{2019ApJ...887L..13D} we noted that our initial three epochs of follow-up comprise the best widefield GHz-frequency transient survey to-date, superseding previous surveys by an order of magnitude in both area and depth. The observations presented in this work has more than tripled the number of epochs, although we note that epoch 6 only has a 74\% overlap with the other pointings. The total areal coverage for this search is $262\,\deg^2$, over four times larger than the initial search which covered 60\,$\deg^2$ (excluding the initial reference epoch in both cases). \citet{2020ApJ...903..116A} report a search for transients with a total areal coverage of $\sim 540\,\deg^2$, but with a detection threshold of $500\,\mu$Jy. Assuming the extragalactic radio transient source count obeys $N\propto S^{-1.5}$ as would be expected in a Euclidean Universe, this makes our observations over twice as sensitive to radio transients those reported by \citet{2020ApJ...903..116A}, although the difference is statistically negligible at such low detection rates\footnote{This estimate ignores the different choices of observing frequency (943\,MHz vs 3\,GHz). However, the spectral index of most radio transients at late times (i.e. typical of their long-term evolution) is generally expected to be negative, making our observations even more sensitive}.

We have detected one radio transient (excluding variable sources such as scintillating AGN) that is likely unrelated to GW190814, ASKAP~J005022.3$-$230349, and therefore measure the surface density of radio transients above $170\,\mu$Jy at 944\,MHz to be $0.0038^{+0.02}_{-0.0037}\,\deg^{-2}$ with uncertainties corresponding to a double-sided 95\% confidence interval \citep[][]{1986ApJ...303..336G}. However, if ASKAP~J005022.3$-$230349 is related to GW190814 then we have detected no transients in our untargeted search and place an upper limit of $0.013\,\deg^2$ on the radio transient surface density for transients above 170\,$\mu$Jy at 95\% confidence.  This measurement is in good agreement with theoretical estimates for the surface density of off-axis long GRBs \citep[][]{2015ApJ...806..224M}, and is also consistent with estimates for tidal disruption events and neutron star mergers.

\subsection{Evaluating follow-up strategies}
While our constraints on the properties of any relativistic afterglow produced by GW190814 are comparable to those of \citet{2021arXiv210208957A}, the observing strategies and resources used to achieve them differ significantly. 

By targeting potential host galaxies within the localisation volume, \citet{2021arXiv210208957A} minimise the total area required to be observed and thereby the total time per-epoch. The smaller, more targeted, areal coverage results in fewer false-positives in general while any transients that are detected are likely associated with the targeted galaxy and therefore fall within the localisation volume of the event. In comparison, our widefield unbiased follow-up predisposes our search to a larger number of false-positives. Some of these can be ruled out via comparison to galaxy catalogues that either classify them as variable AGN or as being associated with galaxies outside of the localisation volume, as we have done in this work. However, many of these require further observations to determine their nature, which could take the form of dedicated follow-up (which is not feasible for tens--hundreds of candidates) or continued widefield monitoring. To ensure that we are left with a manageable number of candidates for human vetting we therefore require either a larger number of observations that will intrinsically decrease the number of false-positives in our sample, or more stringent variability cutoffs. While the latter option may be suitable for widefield untargeted transient searches \citep[e.g.][]{2016ApJ...818..105M}, it is not ideal for gravitational wave follow-up where we know the radio counterparts will likely be faint.

While our widefield many-epoch strategy is more resource intensive, it has a number of advantages to a targeted approach with fewer epochs. The lack of deep all-sky galaxy catalogues means that any galaxy-targeted approach is only feasible for the closest events and simultaneously, the interpretation of any results is always limited by the completeness of the galaxy catalogue. As we move toward the 2030s, galaxy targeting will become a less effective strategy as gravitational wave detector horizons expand, and next-generation radio facilities with higher survey speeds come online \citep[][]{2021MNRAS.505.2647D}.

Observing a small number of epochs necessitates the targeting of the expected peak timescale to maximise the chance of a detection. However, compact object mergers (even when including the population of known short GRBs) are not yet well-understood and therefore this strategy risks limiting the detectable sample to events that fit canonical models. Additionally, observations targeting the peak of the lightcurve may detect radio emission from a merger that does not pass the relevant transient detection thresholds. For example, if the observations occur either side of the lightcurve peak and measure similar flux densities, or occur with insufficient time between them to detect any source evolution.

Future widefield gravitational follow-up observations with next-generation facilities will provide a unique opportunity for serendipitous discoveries. In Section \ref{subsec:rtr} we demonstrated that the observations reported in this work comprise an unparalleled dataset in terms of sensitivity, areal coverage and number of repetitions. \citet[][]{2021MNRAS.502.3294W} discovered a Galactic plasma filament in a search for short timescale variability in these observations, while  \citet[][]{2021MNRAS.505L..11K} have combined the first eight epochs of this search to form the deepest ASKAP observation to-date and found a new extragalactic circular radio source. While the utility of follow-up observations for unrelated science goals is short-lived as they will eventually be superseded by large-scale dedicated surveys, they will still produce useful data in the meantime.

\section{Conclusions}
We have carried out further follow-up observations of the possible neutron star--black hole merger GW190814 with the Australian Square Kilometre Array Pathfinder, building upon the work originally reported by \citet[][]{2019ApJ...887L..13D}. Our ten epochs of observation were carried out on an approximately logarithmic cadence out to 655 days post-merger and cover 30\,$\deg^2$, comprising 87\% of the merger localisation.

We used two techniques to search for a radio counterpart to the merger. We carried out a widefield transient search of the entire field, which found 187 initial candidates that passed our initial search criteria and variability metrics. A qualitative analysis of the images and lightcurve morphology of all 187 sources found that only 12 were real sources with lightcurves resembling those expected from extragalactic synchrotron transients. After a more detailed analysis including comparison to archival multiwavelength data we find that only one candidate is likely to be an intrinsically transient source. However, we are able to rule it out as a counterpart to GW190814 based on its steep, late-time, rise which is incompatible with even the most extreme radio afterglow models. We also carried out a targeted search around known galaxies and found no viable counterparts. In addition, we have also used our observations to comprehensively rule out the candidate counterpart found by \citet[][]{2021arXiv210208957A}.

These observations comprise the most sensitive widefield radio transient survey to-date, and based on our detection of a single transient (likely unrelated to GW190814) we estimate the surface density of radio transients above 170\,$\mu$Jy at 944\,MHz to be $0.0043^{+0.02}_{-0.0042}\,\deg^{-2}$. This survey has helped set expectations for searches for radio transients that are in their early stages \citep[][]{2021arXiv210806039M,2016mks..confE..13F,2020PASP..132c5001L}, as well as those that will be performed with next-generation telescopes like the Square Kilometre Array \citep[][]{2015aska.confE..51F}.

The continued non-detection of a radio counterpart to GW190814 allows us to improve our previous constraints on the circum-merger density, $n$ and merger inclination, $\theta_{\rm obs}$. However, our limits are not sufficiently constraining to confirm that this merger did not produce a radio counterpart. 

The fourth gravitational wave observing run (O4) is expected to begin in mid-2022 after upgrades to Virgo and both LIGO detectors and the Kamioka Gravitational Wave Detector (KAGRA), is also expected to join the run. The improved detector network sensitivity will result in a higher merger detection rate and better localisation capabilities, both of which will lead to a larger number of events for which electromagnetic follow-up is feasible. We expect ASKAP to take a leading role in this effort with a focus on localising events that do not produce a kilonova, or those that are not possible to follow up with optical facilities due to observing constraints.

\section*{Acknowledgements}
We thank Kate Alexander and Andrew Zic for useful discussions.

TM acknowledges the support of the Australian Research Council through grant DP190100561. DK and AO are supported by NSF grant AST-1816492. JL and JP are supported by the Australian Government Research Training Program Scholarship. 
Parts of this research were conducted by the Australian Research Council Centre of Excellence for Gravitational Wave Discovery (OzGrav), project number CE170100004.

The Australian SKA Pathfinder is part of the Australia Telescope National Facility which is managed by CSIRO. Operation of ASKAP is funded by the Australian Government with support from the National Collaborative Research Infrastructure Strategy. ASKAP uses the resources of the Pawsey Supercomputing Centre. Establishment of ASKAP, the Murchison Radio-astronomy Observatory and the Pawsey Supercomputing Centre are initiatives of the Australian Government, with support from the Government of Western Australia and the Science and Industry Endowment Fund. We acknowledge the Wajarri Yamatji people as the traditional owners of the Observatory site.

This work has made use of data from the European Space Agency (ESA) mission
{\it Gaia} (\url{https://www.cosmos.esa.int/gaia}), processed by the {\it Gaia}
Data Processing and Analysis Consortium (DPAC,
\url{https://www.cosmos.esa.int/web/gaia/dpac/consortium}). Funding for the DPAC
has been provided by national institutions, in particular the institutions
participating in the {\it Gaia} Multilateral Agreement.

This publication makes use of data products from the Wide-field Infrared Survey Explorer\citep{2010AJ....140.1868W}, which is a joint project of the University of California, Los Angeles, and the Jet Propulsion Laboratory/California Institute of Technology, funded by the National Aeronautics and Space Administration. 

The SALT observations were obtained under the SALT Large Science Programme on transients (2018-2-LSP-001; PI: DAHB) which is also supported by Poland under grant no. MNiSW DIR/WK/2016/07. DAHB acknowledges research support from the National Research Foundation. MG is supported by the EU Horizon 2020 research and innovation programme under grant agreement No 101004719.

This research has made use of the CIRADA cutout service at URL \url{cutouts.cirada.ca}, operated by the Canadian Initiative for Radio Astronomy Data Analysis (CIRADA). CIRADA is funded by a grant from the Canada Foundation for Innovation 2017 Innovation Fund (Project 35999), as well as by the Provinces of Ontario, British Columbia, Alberta, Manitoba and Quebec, in collaboration with the National Research Council of Canada, the US National Radio Astronomy Observatory and Australia’s Commonwealth Scientific and Industrial Research Organisation.

This research has made use of NASA's Astrophysics Data System Bibliographic Services.

This research has made use of the NASA/IPAC Extragalactic Database (NED) which is operated by the Jet Propulsion Laboratory, California Institute of Technology, under contract with the National Aeronautics and Space Administration. This paper made use of WebPlotDigitizer (\url{http://arohatgi.info/WebPlotDigitizer/}) by Ankit Rohatgi The acknowledgements were compiled using the Astronomy Acknowledgement Generator. This research made use of matplotlib, a Python library for publication quality graphics \citep{Hunter:2007} This research made use of SciPy \citep{Virtanen_2020} This research made use of Astropy, a community-developed core Python package for Astronomy \citep{2018AJ....156..123A, 2013A&A...558A..33A} This research made use of NumPy \citep{harris2020array} This research made use of pandas \citep{McKinney_2010, McKinney_2011}

\section*{Data Availability}
The ASKAP data used in this paper can be accessed through the CSIRO ASKAP Science Data Archive (CASDA\footnote{\url{https://data.csiro.au/dap/public/casda/casdaSearch.zul}}) under project code AS111.

The SALT spectrum of the optical counterpart candidate for ASKAP J005022.3-230349 can be obtained from the SAAO cloud sever at this address: \url{https://cloudcape.saao.ac.za/index.php/s/XzuMSDVZtzLpnBo}



\bsp	
\label{lastpage}
\end{document}

%% file: obs_table.tex
\begin{table}
    \centering
    \begin{tabular}{cccrrc}
    \hline\hline
    Epoch & SBID & Start Time & $\Delta$T & Int. Time & $\sigma_{\rm RMS}$\\
     &  & (UTC) & (days) & (hours) & ($\mu$Jy)\\
    \hline
    1 & 9602 & 2019-08-16 14:11:27 & 2.6 & 10.7 & 35\\
    2 & 9649 & 2019-08-23 13:42:59 & 9.6 & 10.7 & 39\\
    3 & 9910 & 2019-09-16 12:08:34 & 33.5 & 10.6 & 39\\
    4 & 10463 & 2019-11-07 08:44:10 & 85.4 & 10.6 & 39\\
    5 & 12704 & 2020-04-03 23:00:00 & 234 & 15.3 & 28\\
    6 & 13570 & 2020-04-29 21:41:11 & 260 & 10.0& 38\\
    7 & 15191 & 2020-07-03 17:01:21 & 325 & 10.5 & 31\\
    8 & 18925 & 2020-11-28 09:18:30 & 472 & 9.0 & 35\\
    9 & 18912 & 2020-11-29 07:15:31 & 473 & 7.1 & 46\\
    10 & 27379 & 2021-05-29 19:23:44 & 655 &  10.6& 31\\
    \hline\hline
    \end{tabular}
    \caption{ASKAP follow-up observations of GW190814 centered on $\alpha=00^{\rm h}50^{\rm m}37\fs5$, $\delta=-25\degr16\arcmin57\fs37$ (J2000). All observations were carried out with 288\,MHz of bandwidth centered on 943\,MHz. Epoch 6 was rotated by 67.5 degrees with respect to the other observations and centered on $\alpha=00^{\rm h}58^{\rm m}00$, $\delta=-23\degr45\arcmin00$. Data products from all observations are publicly available from the CSIRO ASKAP Science Data Archive under project code AS111 with the Schedulude Block ID (SBID) given in column 2.}
    \label{tab:obs_descrip}
\end{table}